\documentclass[sigconf]{acmart}
\AtBeginDocument{%
  }
\pdfoutput=1

\usepackage{amsmath,amssymb,amsfonts}

\usepackage{totpages}
\usepackage{graphicx}
\usepackage{forest}
\usepackage{subcaption}
\usepackage{textcomp}
\usepackage{tikz}
\usepackage{xcolor}
\usepackage{multirow}
\usepackage{algorithm}
\usepackage{algpseudocode}
\usepackage{pgfplots}
\usepackage{amsmath} 
\pgfplotsset{compat=1.18} 
\usepackage{lipsum} 
\usepackage{booktabs} 
\usepackage{siunitx}  
\usepackage{rotating} 
\usepackage{tabularx}
\usepackage{multirow}
\usepackage{geometry}
\usepackage{listings}
\usepackage{wrapfig}
\usepackage{colortbl}
\usepackage{tcolorbox}
\tcbuselibrary{listings}

\definecolor{codegreen}{rgb}{0,0.6,0}
\definecolor{codegray}{rgb}{0.5,0.5,0.5}
\definecolor{codepurple}{rgb}{0.58,0,0.82}
\definecolor{backcolour}{rgb}{0.95,0.95,0.92}

\lstdefinestyle{mystyle}{
    backgroundcolor=\color{backcolour},
    commentstyle=\color{codegreen},
    keywordstyle=\color{magenta},
    numberstyle=\tiny\color{codegray},
    stringstyle=\color{codepurple},
    basicstyle=\ttfamily\footnotesize,
    breakatwhitespace=false,
    breaklines=true,
    captionpos=b,
    keepspaces=true,
    numbers=left,
    numbersep=5pt,
    showspaces=false,
    showstringspaces=false,
    showtabs=false,
    tabsize=2
}

\setcopyright{acmlicensed}
\copyrightyear{2026}
\acmYear{2026}
\acmDOI{XXXXXXX.XXXXXXX}
\begin{document}

\title{GRACE: Cluster-Specific Sequence Reuse for Compiler Auto-Tuning}

\author{Haolin Pan}
\affiliation{%
  \institution{Hangzhou Institute for Advanced Study at UCAS}
  \city{Hangzhou}
  \country{China}
}
\affiliation{%
  \institution{Institute of Software, Chinese Academy of Sciences}
  \city{Beijing}
  \country{China}
}
\affiliation{%
  \institution{University of Chinese Academy of Sciences}
  \city{Beijing}
  \country{China}
}
\email{panhaolin21@mails.ucas.ac.cn}

\author{Chao Zha}
\affiliation{%
  \institution{Institute of Computing Technology, Chinese Academy of Sciences}
  \city{Beijing}
  \country{China}}
\affiliation{%
  \institution{Research Center for High Efficiency Computing Infrastructure, Zhejiang Lab}
  \city{Zhejiang}
  \country{China}}
\affiliation{%
  \institution{University of Chinese Academy of Sciences}
  \city{Beijing}
  \country{China}}
\email{zhachao21@mails.ucas.ac.cn}

\author{Jinyuan Dong}
\affiliation{%
  \institution{Institute of Software, Chinese Academy of Sciences}
  \city{Beijing}
  \country{China}
}
\affiliation{%
  \institution{University of Chinese Academy of Sciences}
  \city{Beijing}
  \country{China}}
\email{dongjinyuan24@mails.ucas.ac.cn}

\author{Mingjie Xing}
\authornote{Corresponding Author.}
\affiliation{%
  \institution{Institute of Software, Chinese Academy of Sciences}
  \city{Beijing}
  \country{China}}
\email{mingjie@iscas.ac.cn}

\author{Yanjun Wu}
\affiliation{%
  \institution{Institute of Software, Chinese Academy of Sciences}
  \city{Beijing}
  \country{China}}
\affiliation{%
  \institution{University of Chinese Academy of Sciences}
  \city{Beijing}
  \country{China}}
\email{yanjun@iscas.ac.cn}

\renewcommand{\shortauthors}{Haoin Pan et al.}

\begin{abstract}

Compiler auto-tuning aims to improve optimization quality beyond fixed compiler heuristics, but existing approaches often face a trade-off between effectiveness and deployability. Iterative compilation can discover strong program-specific optimization sequences, yet its search cost is often prohibitive for practical reuse. Learning-based methods reduce tuning overhead, but their effectiveness depends on how well optimization knowledge transfers to unseen programs. Recent coreset-based methods improve this trade-off, but they typically either still rely on relatively large test-time search or assume that a single global coreset can serve all programs well. We present GRACE, a compiler auto-tuning framework based on \emph{cluster-specific sequence reuse}. GRACE constructs a small reusable sequence coreset for each group of similar programs by combining global pass synergy analysis, optimization-response-guided program organization, and cluster-specific evolutionary search. At deployment time, it evaluates a small coreset on the target program and optionally performs lightweight refinement within a restricted search space, yielding bounded overhead. We evaluate GRACE on seven benchmark datasets using LLVM 10.0.0 and LLVM 18.1.6. For code-size optimization, GRACE reduces LLVM IR instruction count by 9.92\% and 10.30\% on average relative to \texttt{opt -Oz}, while requiring less than 1\,s tuning time per program at deployment. Under an execution-oriented objective, GRACE reduces estimated cycle counts by 26.84\% and 27.54\% on average relative to \texttt{opt -O3}, and also yields measurable end-to-end speedups on runnable cBench and polybench programs. These results suggest that offline-constructed, cluster-specific sequence coresets provide a practical balance between optimization quality and cost.

\end{abstract}




\maketitle

\section{Introduction}
\label{sec:introduction}
Modern compilers apply sequences of optimization passes to improve program characteristics such as code size, execution performance, and energy efficiency. The relative importance of these objectives varies across application domains and deployment settings. Compiler infrastructures such as LLVM \cite{lattner2004llvm} provide modular pass pipelines and predefined optimization levels (e.g., \texttt{-Oz} and \texttt{-O3}) to balance compilation cost and optimization effectiveness. However, these heuristic pipelines are designed for broad applicability rather than program-specific optimization, and often leave room for improvement because pass selection and phase ordering remain highly complex.

Prior work has attempted to address this problem. Iterative Compilation (IC) methods search the optimization space by evaluating many candidate sequences \cite{GA,opentuner,RIO,tpe,Comptuner,BOCA,ICMC,zhu2025pdcat}. Although such methods can find strong program-specific solutions, they are often too costly for settings that require fast compilation or scalable reuse. Learning-based methods reduce search cost by predicting promising sequences or optimization decisions \cite{autophase,wang2018machine,leather2020machine,cummins2023large,pan2025compiler-r1}, but their effectiveness depends on how well the learned mapping transfers to unseen programs. In practice, it remains difficult to obtain both strong optimization quality and low deployment cost at the same time. For code size optimization, some recent methods have sought to improve this trade-off by exploiting reusable optimization knowledge through compact candidate sets or \emph{coresets}. Representative examples include CFSAT \cite{cfsat} and Coreset-NVP \cite{coreset}. These approaches are attractive because they can reduce exploration cost while still achieving good empirical performance. However, they reuse optimization knowledge in different ways, and each still has limitations. CFSAT reuses synergistic pass relationships to narrow down the optimization space, which makes search more focused; however, its test-time optimization still depends on searching within a relatively large space. Coreset-NVP instead reuses complete optimization sequences and predicts which sequence in a global coreset is likely to work best for a new program. This avoids large online search, but it assumes that one shared global coreset can serve all programs well, which may not always be flexible enough when different program groups prefer different optimization sequences.

In this paper, we present GRACE, a compiler auto-tuning framework based on a different reuse strategy: \emph{cluster-specific sequence reuse}. The key idea is simple: instead of reusing only local pass relationships as in CFSAT, or relying on one global sequence coreset for all programs as in Coreset-NVP, GRACE aims to build a small reusable sequence coreset for each group of similar programs. Concretely, GRACE first performs global pass synergy analysis to identify an empirically promising candidate space and a constrained pass pool. It then organizes training programs using representations guided by optimization-response similarity and partitions them into clusters. Within each cluster, GRACE performs evolutionary search to derive representative pass sequences, producing a compact coreset of reusable candidates. At deployment time, GRACE does not attempt to predict a single best sequence directly. Instead, it evaluates a small cluster-specific coreset on the unseen program and optionally applies lightweight refinement within a restricted search space. In this way, GRACE still performs test-time search, but keeps the search cost explicitly bounded and substantially smaller than searching over a much larger candidate space.

We evaluate GRACE on seven benchmark datasets using LLVM 10.0.0 and LLVM 18.1.6. For code size optimization, GRACE reduces LLVM IR instruction count by an average of \textbf{9.92\%} on LLVM 10.0.0 and \textbf{10.30\%} on LLVM 18.1.6 relative to \texttt{opt -Oz}, while requiring less than \textbf{1s} tuning time per program at deployment. We further study the same framework under an execution-oriented objective. In that setting, GRACE identifies pass sequences that reduce estimated cycle counts by \textbf{26.84\%} and \textbf{27.54\%} relative to \texttt{opt -O3} on the two LLVM versions, and also yields measurable wall-clock speedups on runnable cBench and polybench programs across two hardware platforms. These results suggest that offline-constructed, cluster-specific sequence coresets provide a practical way to balance optimization quality and deployment cost.

\noindent \textbf{Contributions.}
This paper makes three contributions:
\begin{itemize}
    \item We propose GRACE, the first compiler auto-tuning framework to introduce \emph{cluster-specific sequence reuse}. Existing reuse strategies mainly operate at either the level of pair-level search-space reuse or global sequence reuse. GRACE extends sequence reuse to the cluster-specific level by constructing compact reusable sequence coresets for different program clusters.
    \item To realize cluster-specific reuse, we design a \emph{hierarchical sequence discovery framework}. GRACE first extracts optimization priors shared across programs through global pass synergy, then organizes programs using representations guided by optimization-response similarity, and finally performs sequence-level evolution within each program cluster to identify representative reusable sequences.
    \item GRACE shifts test-time decision from relying on a single prediction of the best sequence to controlled empirical evaluation over a small cluster-specific coreset. This reduces dependence on prediction accuracy and enables more robust optimization under bounded deployment cost. We conduct extensive experiments on seven benchmark datasets and two LLVM versions, showing that GRACE achieves better average performance than other baselines while maintaining low deployment overhead.
\end{itemize}
\section{Related Work}
\label{sec:related_work}

Compiler auto-tuning has been studied extensively, with prior work mainly differing in how it balances optimization quality against tuning and deployment cost. Existing approaches can be broadly grouped into iterative search methods, learning-based methods, and sequence-reuse methods.

Iterative Compilation (IC) methods directly explore optimization sequences by repeatedly compiling and evaluating candidate configurations \cite{GA,opentuner,RIO,tpe,Comptuner,BOCA,ICMC,zhu2025pdcat,gao2025grouptuner,lenfers2026schedgehammer,srtuner}. Because they rely on empirical feedback from the target program, they can often find strong program-specific solutions, but their search cost is typically high. Learning-based methods aim to reduce tuning cost by learning mappings from program features to optimization decisions \cite{autophase,wang2018machine,leather2020machine,cfsat}. These approaches improve deployment efficiency, but their effectiveness depends on how well the learned mapping transfers to unseen programs. Sequence-reuse methods further explore this direction by constructing compact candidate sets or reusable sequence pools for deployment, as in CFSAT and Coreset-NVP \cite{cfsat,coreset}. More recently, large language model (LLM)-based methods have also been explored for compiler optimization \cite{pan2025compiler-r1,lin2025awarecompiler,tang2025compiler,copet2025cwm,tang2025compiler,cui2025decos,chen2026magellan}. While they offer a different route to optimization guidance, they often introduce higher inference cost and greater integration complexity in practical compilation pipelines. GRACE is most closely related to the sequence-reuse line, but differs in how reusable sequences are constructed and deployed.

Among the most closely related works, CFSAT and Coreset-NVP represent two different reuse strategies. CFSAT reuses synergistic pass relationships to narrow the optimization space, but still relies on searching within a relatively large space at test time. Coreset-NVP instead reuses complete optimization sequences and predicts which sequence in a global coreset should work best for a new program, thereby avoiding large online search. GRACE differs from both: rather than reusing only local pass relationships as search-space guidance, or relying on a single global sequence coreset, it explores \emph{cluster-specific sequence reuse} by constructing small reusable sequence coresets for different program clusters.
\section{The GRACE Framework}
\label{sec:grace_framework}

\begin{figure*}[htbp]
\centering
    \includegraphics[width=0.95\textwidth]{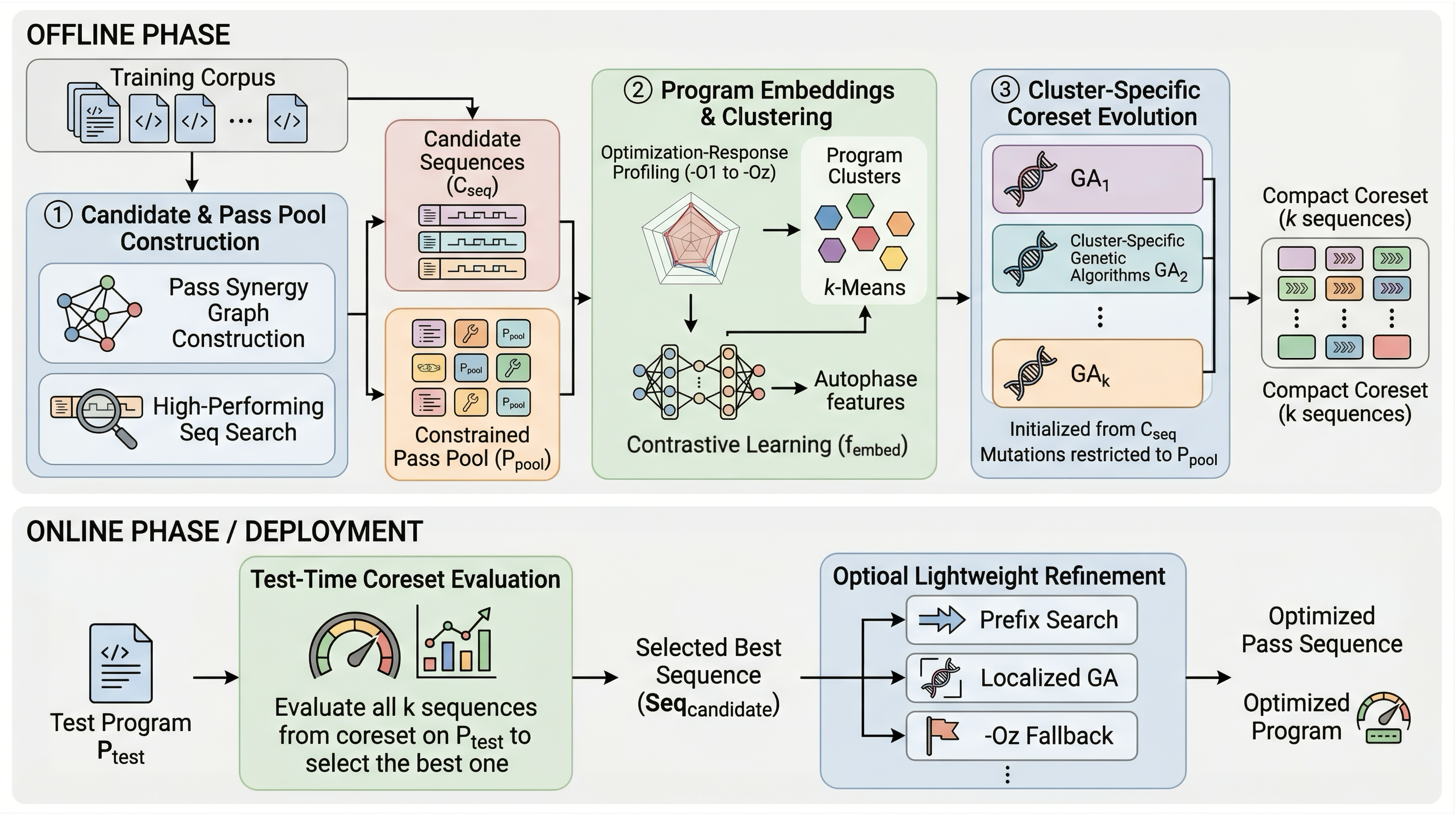}
    \caption{The GRACE framework.} 
    \label{fig:grace_overview}
\end{figure*}

GRACE is designed to move most sequence discovery effort to an offline phase, while keeping deployment-time tuning explicitly bounded. The framework follows four stages. First, it extracts globally shared optimization priors from the training corpus and uses them to construct an empirically promising candidate set together with a constrained pass pool. Second, it learns program representations for offline organization and partitions training programs into clusters. Third, it performs cluster-specific evolutionary search to derive a compact coreset of representative sequences for each cluster. Finally, at test time, GRACE evaluates this precomputed coreset on an unseen program and may optionally apply lightweight refinement within a restricted search space. Figure~\ref{fig:grace_overview} gives an overview of the framework.


\subsection{Candidate and Pass Pool Construction}
\label{ssec:stage1_global_candidates}

\textbf{Rationale for Candidate Construction.} 
The first stage of GRACE aims to extract globally useful optimization priors from the training set $\mathcal{P}_{train}$. Rather than starting later cluster-specific search from the full pass space, GRACE first constructs (i) a set of empirically promising candidate sequences and (ii) a constrained pool of useful passes. These two artifacts serve as global guidance for the later cluster-specific sequence discovery process, biasing search toward regions of the space that already demonstrate favorable optimization outcomes on the training corpus.

\textbf{Pass Synergy Identification and Graph Construction.}
Following prior observations that useful compiler optimizations often arise from interactions between passes \cite{cfsat}, GRACE begins by identifying synergistic pass pairs on the training corpus. For each training program $P_i \in \mathcal{P}_{train}$, we begin by identifying its set of synergistic pass pairs $S_i$. We first compute a baseline IR instruction count $V_P$ for $P_i$. Then, for each compiler pass $B \in \mathcal{O}_{all}$, we evaluate its effect on $P_i$. If applying $B$ alone reduces the instruction count, we further explore its synergy with every other pass $A \in \mathcal{O}_{all}$ by applying the sequence $\langle A, B \rangle$ to the original $P_i$. If this combination yields a further reduction in instruction count compared to $B$ alone ($V_{P_{AB}} < V_{P_B}$), the pair $(A, B)$ is added to $S_i$. After processing all programs, we aggregate the individual synergy sets $\{S_i\}$ into a global collection $\mathcal{S}_{global}$ and build a pass-synergy graph $G_{co}$. This graph serves as a compact representation of pass interactions that recur across the training corpus, and provides global guidance for subsequent sequence construction.

\textbf{High-Performing Sequence Search and Pass Pool Formulation.}
GRACE uses the pass-synergy graph $G_{co}$ to guide the construction of high-performing sequences on the training corpus. We focus on \textbf{pairwise synergy} as a tractable approximation of pass interaction: it captures useful local dependencies while keeping graph construction and subsequent search manageable. Although higher-order interactions may also exist, modeling them explicitly would substantially increase construction cost and search complexity. Leveraging this graph, a search algorithm (e.g., GA) is applied to each training program $P_i$ to identify a high-performing sequence $Seq_{hp,i}$ with low IR instruction count. The resulting unique sequences form the set $Seq_{HP}$.
The resulting unique sequences form the set $Seq_{HP}$. Since GRACE ultimately aims to discover reusable sequences rather than sequences that work well only for a single program, we further evaluate each $Seq_u \in Seq_{HP}$ over the training corpus and assign it a weighted score. The scoring function is defined as:
\begin{equation}
S = w_{avg} \cdot \Delta_{avg} - w_{std} \cdot \sigma - w_{neg} \cdot R_{neg}
\end{equation}
where $\Delta_{avg}$ represents the average improvement ratio, $\sigma$ denotes the standard deviation (reflecting stability), and $R_{neg}$ measures the negative optimization rate (the proportion of programs where the sequence underperforms relative to baseline). The weights are hyper-parameters such that $w_{avg}(0.75) + w_{std}(0.1) + w_{neg}(0.15) = 1$, and were chosen empirically to balance average performance against volatility and negative outcomes. Based on these scores, the top-$k_{top}$ sequences form the initial candidate set $C_{seq}$. Finally, all unique compiler passes appearing in $C_{seq}$ are aggregated into a constrained pass pool $P_{pool}$, which serves as the search basis for the subsequent cluster-specific evolutionary stage.


\begin{figure}[htbp]
\centering
\begin{minipage}{0.5\textwidth}
    \centering
    \includegraphics[width=\textwidth]{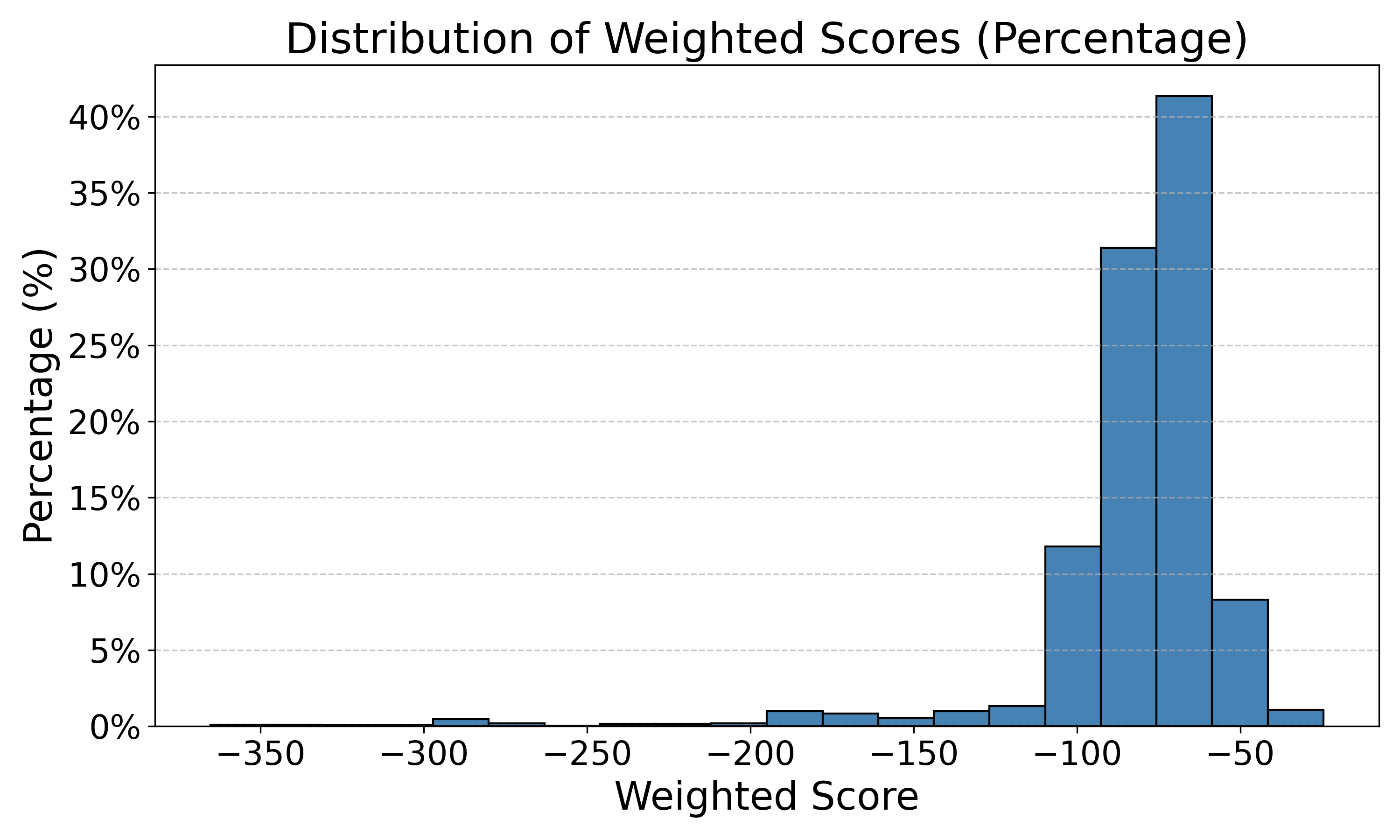}
    \caption{Distribution of weighted scores for unique high-performing sequences. The spread reflects variation in empirical quality across candidate sequences.}
    \label{fig:weighted_score_hist}
\end{minipage}
\end{figure}

\subsection{Program Embeddings and Clustering}
\label{ssec:stage2_contrastive_learning}

\textbf{Rationale for Program Representation and Clustering.}
The second stage of GRACE organizes training programs before cluster-specific sequence discovery. Since GRACE aims to construct reusable sequence coresets for different program groups, it requires an offline structure that separates programs into clusters rather than treating the entire training corpus as a single homogeneous set. In the design of GRACE, this organization is guided by \emph{optimization-response similarity}: programs that respond similarly to a small set of standard compiler optimization levels are encouraged to be placed closer together in the learned representation space. The goal of this stage is not to predict optimization sequences directly, but to provide a more structured basis for the subsequent cluster-specific evolutionary search.

\textbf{Program Representation and Optimization-Response-Guided Positive Sample Generation.}
To represent each training program $P_i$, we begin with Autophase features \cite{autophase}, which capture static program characteristics relevant to compiler optimization. To construct supervision for contrastive learning, we additionally derive an \emph{optimization-response profile} for each training program. Concretely, for every program we apply a fixed set of standard LLVM optimization levels, namely \texttt{-O1}, \texttt{-O2}, \texttt{-O3}, \texttt{-Os}, and \texttt{-Oz}, and record the percentage reduction in instruction count relative to the unoptimized baseline. This yields a 5-dimensional response vector for each program.

We choose these optimization levels because they are canonical compiler configurations that are fixed, widely used, and empirically reflect different optimization preferences. In particular, \texttt{-O1}, \texttt{-O2}, and \texttt{-O3} capture progressively stronger general-purpose optimization settings, while \texttt{-Os} and \texttt{-Oz} emphasize code-size reduction. Compared with using randomly sampled pass sequences, these standard optimization levels provide a more stable and reproducible set of reference probes for characterizing how a program responds to different optimization strategies, while keeping the profiling cost manageable. In this sense, the resulting response vector serves as a compact summary of each program's optimization profile under a small but diverse set of canonical compiler settings.

Given an anchor program $P_i$, we compare its response vector with those of other training programs using cosine similarity. The most similar program is selected as the positive sample for $P_i$, while the remaining programs in the sampled set are treated as negatives. In this way, positive pairs are not formed from syntactic perturbations of the same program, but from different programs that exhibit similar optimization responses under standard compiler settings. This construction encourages the learned representation to group programs with similar response patterns rather than relying only on superficial structural similarity.

\textbf{Contrastive Learning for Program Embeddings.}
Let $x_i$ denote the Autophase feature vector of program $P_i$. GRACE learns a program embedding function $f_{embed}(\cdot)$ through contrastive learning. We use an encoder network $f_{enc}(\cdot)$ to map $x_i$ to an intermediate representation $h_i = f_{enc}(x_i)$, followed by a projection head $f_{proj}(\cdot)$ that produces a projected vector $z_i = f_{proj}(h_i)$ used during contrastive training.

Given a batch of $N$ anchor programs and their corresponding positive samples, we obtain $2N$ projected vectors $\{z_k\}_{k=1}^{2N}$. We adopt the normalized temperature-scaled cross-entropy objective (NT-Xent). After $\ell_2$ normalization, pairwise similarities are computed using cosine similarity between projected vectors. For each sample, its matched positive counterpart is treated as the target, while all other samples in the batch serve as negatives. Formally, let $z_a$ denote a sample and $z_{pos(a)}$ its positive counterpart. The contrastive loss is defined as
\begin{equation}
\label{eq:contrastive_loss_ce}
\mathcal{L}_a = -\log \frac{\exp(\mathrm{sim}(z_a,z_{pos(a)})/\tau)}{\sum_{k=1,k \neq a}^{2N} \exp(\mathrm{sim}(z_a,z_k)/\tau)},
\end{equation}
where $\mathrm{sim}(\cdot,\cdot)$ denotes cosine similarity and $\tau$ is a temperature hyperparameter. The total loss is the average of $\mathcal{L}_a$ over all $2N$ samples in the batch. After training, we discard the projection head and use the encoder $f_{enc}(\cdot)$ as the final embedding function $f_{embed}(\cdot)$, yielding the program embedding $e_i = f_{embed}(P_i)$ for downstream clustering.

\textbf{K-Means Clustering on Program Embeddings.}
Once embeddings $\{e_i\}_{i=1}^{N_{train}}$ have been obtained for all $N_{train}$ training programs, GRACE partitions them into clusters using k-means \cite{K-means,zha2025nids}. The goal is to organize the training corpus into a set of groups that can support later cluster-specific sequence discovery. Specifically, k-means partitions the embeddings into $k$ disjoint clusters $C = \{C_1, C_2, \dots, C_k\}$, where each cluster $C_j$ is represented by a centroid $\mu_j$. The clustering objective is
\begin{equation}
\label{eq:kmeans_objective}
\underset{C}{\text{argmin}} \sum_{j=1}^{k} \sum_{e_i \in C_j} ||e_i - \mu_j||_2^2.
\end{equation}
This yields $k$ groups of training programs in the learned embedding space, which are then used as the basis for the cluster-specific search stage.

A key design choice in GRACE is that these embeddings are used only for \emph{offline} program organization, rather than for making a single test-time routing decision. After clustering, each cluster is used to derive representative reusable sequences, and deployment evaluates the resulting coreset empirically on the target program. This keeps sequence selection explicit at test time and bounds deployment cost by the coreset size, while avoiding reliance on a single predicted sequence for unseen programs.



\subsection{Cluster-Specific Coreset Evolution}
\label{ssec:stage3_coreset_evolution}

Once the training programs have been partitioned into $k$ clusters, the next stage of GRACE derives one representative pass sequence for each cluster, yielding a coreset of $k$ reusable sequences $\{Seq_{coreset,1}, \dots, Seq_{coreset,k}\}$. The goal of this stage is to specialize sequence discovery to each cluster while still benefiting from the global priors extracted in Stage~1.

To achieve this, GRACE runs a separate genetic algorithm (GA) for each cluster $Cluster_j$. A key design choice is that this cluster-specific search is not started from scratch. Instead, it is both initialized and constrained by the artifacts produced in Section~\ref{ssec:stage1_global_candidates}. In particular, the initial population for the GA of $Cluster_j$ is partially seeded from the candidate set $C_{seq}$, so that search begins from sequences that already showed good empirical behavior on the training corpus. In addition, mutation is restricted to passes drawn from the pass pool $P_{pool}$, which keeps the search focused on passes that appeared useful during the global candidate-construction stage.

The fitness of a sequence for $Cluster_j$ is defined by applying that sequence to all programs in $Cluster_j$ and measuring its aggregate performance on the cluster. In this work, the optimization objective is to maximize the weighted score over the cluster. The GA then evolves the population to identify a sequence $Seq_{coreset,j}$ that performs well for the programs assigned to $Cluster_j$. Repeating this process for all $k$ clusters yields a compact coreset of cluster-specific representative sequences.

The resulting coreset is intended to capture reusable optimization strategies at the cluster level rather than through a single global sequence. To summarize the behavior of the evolved sequences, we compute for each coreset sequence its average instruction-count reduction (Avg), the standard deviation of these reductions (Std), and the negative impact rate (NegRate) over the programs in its assigned cluster. 


\subsection{Test-Time Application with Refinement}
\label{ssec:stage4_test_time}

At deployment time, given an unseen program $P_{test}$, GRACE evaluates all $k$ coreset sequences $\{Seq_{coreset,j}\}_{j=1}^{k}$ obtained in Section~\ref{ssec:stage3_coreset_evolution}. The best-performing sequence on $P_{test}$ is selected as the initial candidate, denoted $Seq_{candidate}$. This introduces a fixed test-time cost proportional to the coreset size, making deployment overhead explicit and bounded.

Starting from $Seq_{candidate}$, GRACE may optionally apply lightweight refinement within a restricted search space. Specifically, it can (i) perform derivative search by testing whether a shorter prefix of the sequence outperforms the full candidate, (ii) apply a localized genetic algorithm restricted to the unique passes already appearing in $Seq_{candidate}$ to explore small local improvements, or (iii) fall back to LLVM's standard \texttt{-Oz} pipeline if the evolved sequence does not outperform the baseline.  These refinement steps are optional. In the default deployment setting, GRACE can be applied using fixed coreset evaluation alone, while refinement provides an additional quality--latency trade-off when needed.
\section{Experimental Setup}
\label{sec:experimental_setup}

\subsection{Compiler Environment}
\label{ssec:compiler_env}

All experiments are conducted in the LLVM compiler infrastructure. We evaluate GRACE on two LLVM versions. LLVM~10.0.0 is used for comparability with prior baselines such as CFSAT~\cite{cfsat} and Coreset-NVP~\cite{coreset}, which were originally evaluated on this version or closely related releases. LLVM~18.1.6 is further included to examine the same framework on a more recent compiler release with updated optimization passes and heuristics. All experiments are run on an AMD EPYC 7763 64-Core system, with GRACE interacting with LLVM's \texttt{opt} tool to apply and evaluate optimization sequences.

\subsection{Datasets}
\label{ssec:datasets}

\begin{table}[htbp]
\centering
\caption{Dataset composition detailing the number of programs used for training and testing.}
\label{tab:dataset_composition}
\setlength{\tabcolsep}{6pt}
\begin{tabular}{l l r r}
\toprule
\textbf{Type} & \textbf{Dataset} & \textbf{Train} & \textbf{Test} \\
\midrule
\multirow{6}{*}{Uncurated}
 & blas-v0 & 133 & 29 \\
 & github-v0 & 7{,}000 & 0 \\
 & linux-v0 & 4{,}906 & 0 \\
 & opencv-v0 \cite{opencv} & 149 & 32 \\
 & poj104-v1 \cite{poj} & 7{,}000 & 0 \\
 & tensorflow-v0 \cite{tensorflow} & 415 & 90 \\
\midrule
\multirow{4}{*}{Curated}
 & cbench-v1 \cite{cbench} & 0 & 11 \\
 & mibench-v1 \cite{mibench} & 0 & 40 \\
 & chstone-v0 \cite{chstone} & 0 & 12 \\
 & npb-v0 \cite{npb} & 0 & 121 \\
\midrule
\textbf{Total} & -- & \textbf{19{,}603} & \textbf{335} \\
\bottomrule
\end{tabular}

\end{table}

We use benchmark datasets from CompilerGym~\cite{CompilerGym} and follow the train--test split protocol of CFSAT~\cite{cfsat}. Table~\ref{tab:dataset_composition} summarizes the resulting dataset composition. The training set is dominated by large-scale uncurated repositories, while the test set includes both held-out programs from overlapping domains and curated benchmark suites from distinct domains. This setup allows us to evaluate whether offline-constructed reusable sequence coresets transfer beyond the programs used during training. To reduce the risk of near-duplicate leakage, we additionally perform a structural similarity scan across the two splits and find no training--testing pair with similarity $\ge 80\%$.

\begin{table*}[htbp]
\centering
\caption{Main code-size optimization results on LLVM 10.0.0. The table reports average LLVM IR instruction count reduction (\%) relative to \texttt{-Oz} and deployment-time cost (s) per program. Higher values indicate greater code-size reduction. Bold indicates the best value in each column.}
\label{tab:main_results_overoz}
\sisetup{table-align-text-post=false} 
\resizebox{\textwidth}{!}{%
\begin{tabular}{@{}l | S[table-format=-2.2] S[table-format=-2.2] S[table-format=-2.2] S[table-format=-2.2] S[table-format=2.2] S[table-format=-1.2] S[table-format=1.2] S[table-format=-2.2] r@{}}
\toprule
\textbf{Method} & {\textbf{blas}} & {\textbf{cbench}} & {\textbf{chstone}} & {\textbf{mibench}} & {\textbf{npb}} & {\textbf{opencv}} & {\textbf{tensorflow}} & {\textbf{Avg.}} & {\textbf{Time (s)}} \\ 
\midrule
\multicolumn{10}{l}{\textit{\textbf{Iterative Search Baselines}}} \\
\midrule
Opentuner \cite{opentuner} &  1.60 &  1.99 & 6.46 & 3.33 & 26.19 & 1.76 & 1.29 &  6.09 & 200 \\
PDCAT \cite{zhu2025pdcat}  & -1.56 &  1.54 & 6.06 & 3.87 & 25.56 & 1.53 & 0.87 &  5.41 & 3600 \\
GA \cite{GA}               & -1.91 &  1.99 & 6.51 & 0.90 & 25.63 & 1.76 & 1.29 &  5.17 & 593 \\
TPE \cite{tpe}             & -2.24 &  0.97 & 7.60 & 0.20 & 24.62 & 1.46 & 1.23 &  4.83 & 905 \\
RIO \cite{RIO}             & -2.02 &  0.24 & 4.98 & 3.47 & 23.87 & 0.79 & 1.23 &  4.65 & 200 \\
BOCA \cite{BOCA}           & -2.36 & -0.16 & 3.18 & -0.69& 22.87 & 1.13 & 1.22 &  3.60 & 3310 \\
CompTuner \cite{Comptuner} & -3.06 & -0.65 & 4.38 & -0.45& 22.99 & 0.44 & 1.01 &  3.52 & 10800 \\

\midrule
\multicolumn{10}{l}{\textit{\textbf{Sequence-Level / Learning-Based Baselines}}} \\
\midrule
CFSAT \cite{cfsat}         & 4.60  &\textbf{5.80}  & \textbf{11.90} & 3.70 & 25.90 & \textbf{5.90} & 6.10 & 9.13 & 6 \\
Coreset-NVP \cite{coreset} & 2.60  & 3.50  & 9.30  & 1.70 & 9.80  & 5.20 & 6.10 & 5.46 & 11 \\
Autophase (PPO-LSTM) \cite{autophase} & -1.12 & 5.60 & 4.49 & 4.41 & -4.67 & -0.09 & 0.05 & 1.24 & 3 \\
Autophase (PPO-noLSTM)     & -4.77 & -79.69 & -80.90 & -107.33 & -76.69 & -2.32 & -0.76 & -50.35 & 2 \\
\rowcolor{gray!10} 
\textbf{GRACE (Ours)}      & {\textbf{5.35}} & {5.65} & {11.32} & {\textbf{7.11}} & {\textbf{27.34}} & {5.67} & {\textbf{6.99}} & {\textbf{9.92}} & \textbf{<1} \\
\bottomrule
\end{tabular}%
}
\end{table*}

\subsection{Baseline Methods}
\label{ssec:baselines}

We compare GRACE with a set of established baselines, including standard compiler heuristics, iterative search methods, and learning-based sequence optimization methods. To improve comparability, all baselines are evaluated within the same LLVM environment and under the same objective used in each experiment (LLVM IR instruction count reduction or cycle minimization).

The baselines include LLVM's standard optimization pipelines (\texttt{-Oz} and \texttt{-O3}), iterative search methods such as GA~\cite{GA}, TPE~\cite{tpe}, RIO~\cite{RIO}, OpenTuner~\cite{opentuner}, CompTuner~\cite{Comptuner}, BOCA~\cite{BOCA}, and PDCAT~\cite{zhu2025pdcat}, as well as learning-based or reusable-sequence methods including Autophase~\cite{autophase}, Coreset-NVP~\cite{coreset}, and CFSAT~\cite{cfsat}. Among these baselines, CFSAT and Coreset-NVP are the most closely related to GRACE: CFSAT reuses synergistic pass relationships to guide search in a reduced space, whereas Coreset-NVP reuses complete optimization sequences through a global coreset. These two methods therefore provide the most direct points of comparison for evaluating GRACE's cluster-specific sequence reuse strategy.

\section{Results and Analysis}
\label{sec:results}

\subsection{Evaluation Metrics}

We evaluate GRACE against established baselines using two primary metrics. For code-size optimization, we report \textit{OverOz}, defined as
$(N_{Oz} - N_{GRACE}) / N_{Oz} \times 100\%$,
where $N_{Oz}$ denotes the LLVM IR instruction count after applying \texttt{opt -Oz}. LLVM IR instruction count provides a stable and reproducible proxy for code-size-oriented optimization in large-scale evaluation, and has been used in prior work under similar settings \cite{seeker2024revealing,coreset}. A higher \textit{OverOz} score indicates a greater reduction in IR instruction count relative to \texttt{-Oz}. For execution-oriented optimization, we primarily report the reduction in predicted execution cycles relative to \texttt{opt -O3}, estimated using the LLVM Machine Code Analyzer (\texttt{llvm-mca}). Similar to IR instruction count for code-size evaluation, this metric provides a stable and reproducible basis for comparing methods under the same protocol. We additionally report end-to-end wall-clock runtime on cBench programs to assess practical speedups.
Unless otherwise specified, GRACE results are reported with a coreset size of $k=100$.

\begin{figure*}[htbp]
    \centering
    \includegraphics[width=0.98\textwidth]{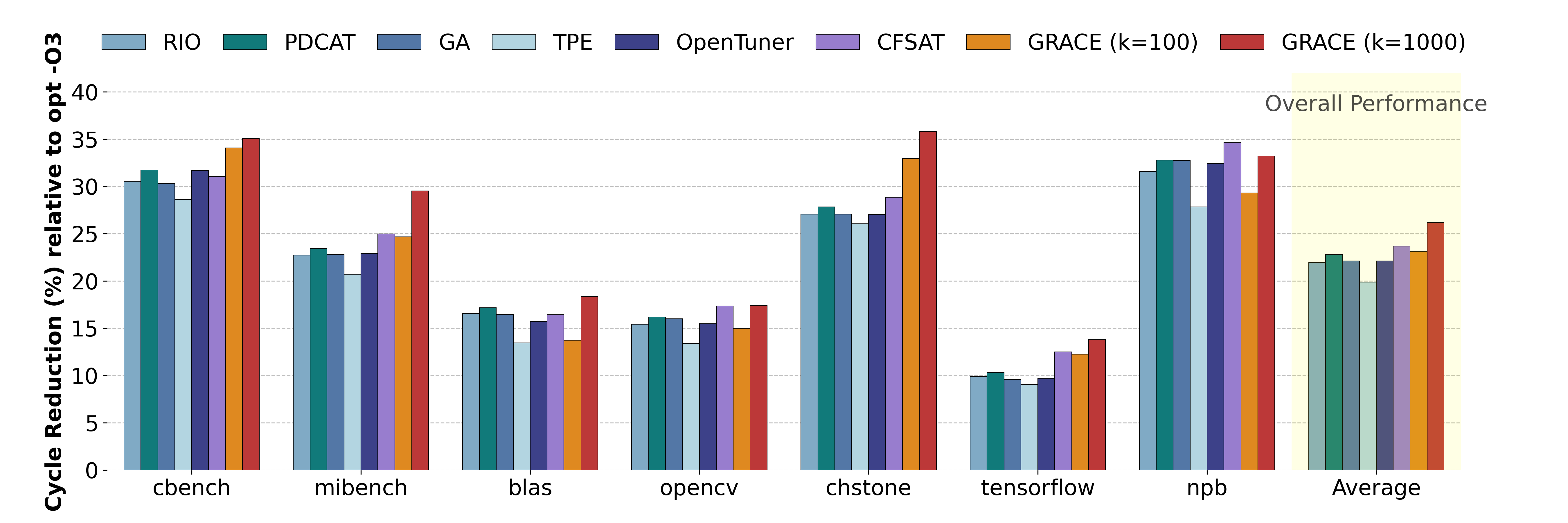} 
    \caption{Average predicted cycle reduction (\%) relative to \texttt{opt -O3} on LLVM 10.0.0. Higher values indicate better execution-oriented optimization. GRACE is shown with two coreset sizes ($k=100$ and $k=1000$).}
    \label{fig:cycles_comparison_bar}
\end{figure*}

\subsection{Code Size (IR Instruction Count)}

Table~\ref{tab:main_results_overoz} reports the main code-size optimization results in terms of LLVM IR instruction count reduction relative to \texttt{opt -Oz}. GRACE achieves an average \textit{OverOz} of \textbf{10.03\%} across the seven datasets, and obtains the highest dataset-level reduction on five of them: \texttt{blas} (5.35\%), \texttt{cbench} (6.45\%), \texttt{mibench} (7.11\%), \texttt{npb} (27.34\%), and \texttt{tensorflow} (6.99\%). At the same time, its deployment-time cost remains below \textbf{1 second} per program on average.

Compared with iterative search baselines, GRACE achieves substantially better average code-size reduction while requiring much lower deployment cost. This suggests that offline-constructed reusable sequence coresets can retain much of the benefit of empirical search without incurring the high online tuning overhead of iterative methods. Among the most closely related sequence-level baselines, GRACE and CFSAT achieve the strongest overall results. CFSAT is better on \texttt{chstone} (11.90\%) and \texttt{opencv} (5.90\%), while GRACE achieves the better average result across datasets (10.03\% vs.\ 9.13\%). A possible explanation is that the two methods make different deployment-time trade-offs: CFSAT continues search in a larger predicted space, whereas GRACE evaluates a fixed coreset to keep test-time cost explicitly bounded. Compared with Coreset-NVP, GRACE also achieves clearly stronger average reduction (10.03\% vs.\ 5.46\%), which suggests that constructing cluster-specific sequence coresets is more effective in this setting than relying on a single global coreset for all programs.

\subsection{Execution Performance (llvm-mca Cycles)}

We also evaluate GRACE under an execution-oriented objective by replacing code-size reduction with predicted cycle reduction relative to \texttt{opt -O3}. Figure~\ref{fig:cycles_comparison_bar} summarizes the results across benchmarks. Under the default configuration ($k=100$), GRACE achieves an average cycle reduction of \textbf{23.62\%}, which is competitive with CFSAT (23.70\%) and PDCAT (22.79\%). When the coreset size is increased to $k=1000$ (denoted GRACE$_{1000}$), the average reduction improves further to \textbf{26.84\%}, exceeding the compared baselines in this setting. GRACE$_{1000}$ also outperforms CFSAT on a majority of programs in our paired comparison, while maintaining an average tuning time of approximately 2 seconds per program.

These results suggest that the GRACE framework is not limited to code-size optimization. The comparison between GRACE$_{100}$ and GRACE$_{1000}$ also illustrates a clear quality--cost trade-off controlled by the coreset size: a smaller coreset provides lower deployment cost, while a larger coreset improves optimization quality by allowing broader bounded evaluation at test time. This behavior is consistent with the design of GRACE, which shifts most sequence discovery effort offline and leaves a tunable amount of empirical evaluation for deployment.

\subsection{End-to-End Execution Time Analysis}
\label{sssec:end_to_end_time}

\begin{table}[t]
\centering
\caption{End-to-end wall-clock runtime normalized to \texttt{-O3} on Intel Xeon 8575C and AMD EPYC 7763 for runnable cBench programs in our environment.}
\begin{tabular}{lcc|lcc}
\toprule
Benchmark & Intel & AMD & Benchmark & Intel & AMD \\
\midrule
dijkstra & 1.073 & 1.056 & tiffmedian & 8.668 & 1.061 \\
patricia & 1.284 & 1.032 & jpeg\_c & 4.949 & 1.025 \\
qsort1 & 1.062 & 1.012 & CRC32 & 1.139 & 1.075 \\
rsynth & 1.203 & 1.378 & tiffdither & 8.678 & 1.101 \\
stringsearch1 & 1.000 & 1.000 & tiff2bw & 12.037 & 1.085 \\
susan\_s & 1.265 & 1.116 & adpcm\_c & 2.159 & 1.921 \\
bzip2e & 1.204 & 1.099 & jpeg\_d & 19.250 & 1.183 \\
bitcount & 1.248 & 1.000 & adpcm\_d & 1.236 & 1.150 \\
susan\_e & 1.000 & 1.000 & tiff2rgba & 12.206 & 1.143 \\
bzip2d & 1.226 & 1.068 & gsm & 1.668 & 1.000 \\
blowfish\_d & 1.472 & 1.078 & blowfish\_e & 1.210 & 1.040 \\
sha & 1.243 & 1.077 & rijndael\_d & 41.389 & 1.011 \\
susan\_c & 1.648 & 1.233 & rijndael\_e & 50.103 & 1.001 \\
\midrule
\multicolumn{6}{c}{\textbf{Geomean Speedup:} Intel = \textbf{2.7$\times$}, AMD = \textbf{1.1$\times$}} \\
\bottomrule
\label{tab:cbench_runtime}
\end{tabular}
\end{table}

\begin{figure}[t]
    \vspace{-1.0em}
    \centering
    \includegraphics[width=0.98\linewidth]{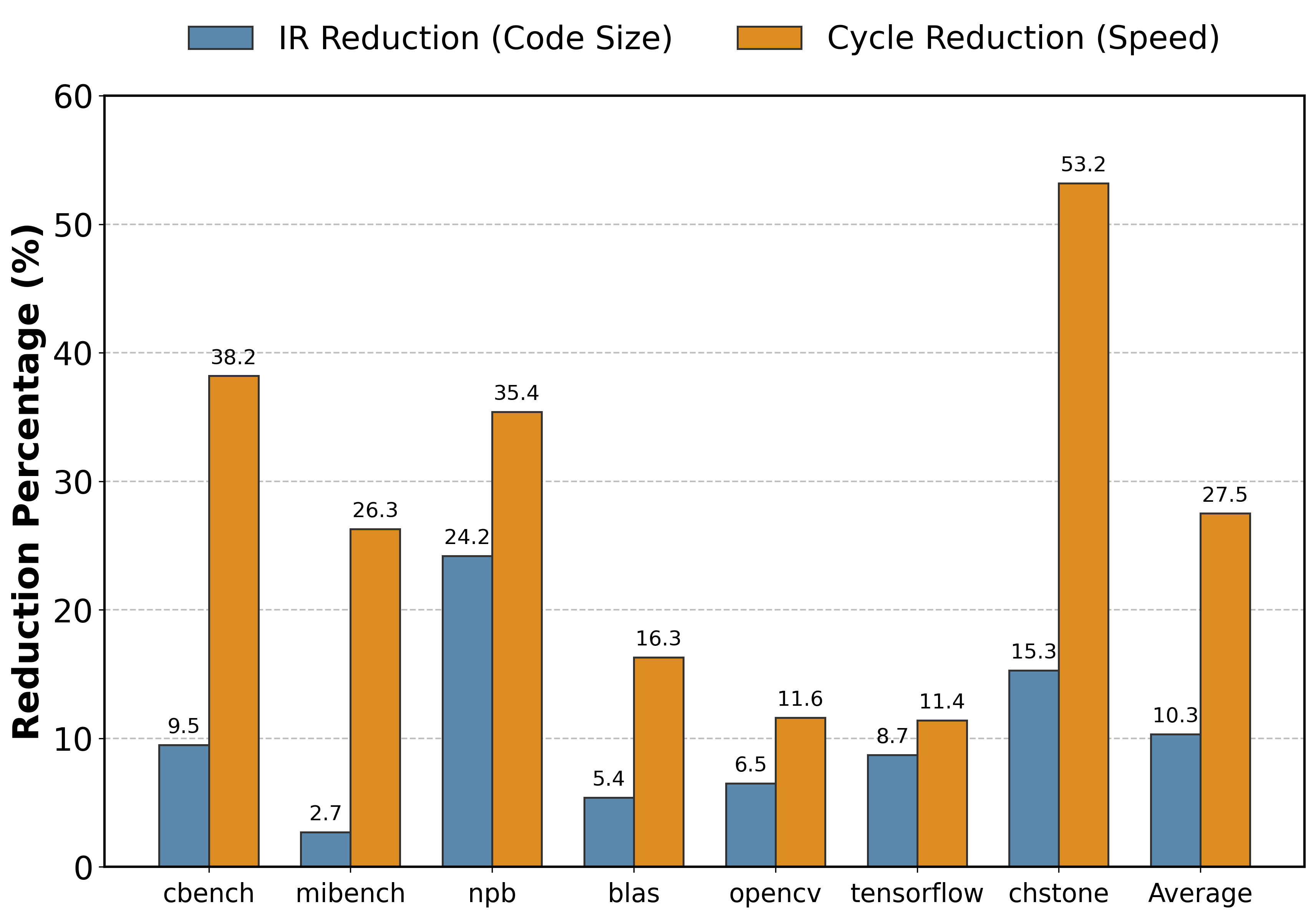} 
    \caption{Performance of GRACE on LLVM 18.1.6, reported as IR instruction count reduction relative to \texttt{opt -Oz} and predicted cycle reduction relative to \texttt{opt -O3} across seven datasets.}
    \label{fig:llvm18_performance}
    \vspace{-1.0em}
\end{figure}

\begin{figure*}[t]
    \vspace{-1.0em}
    \centering
    \includegraphics[width=0.98\linewidth]{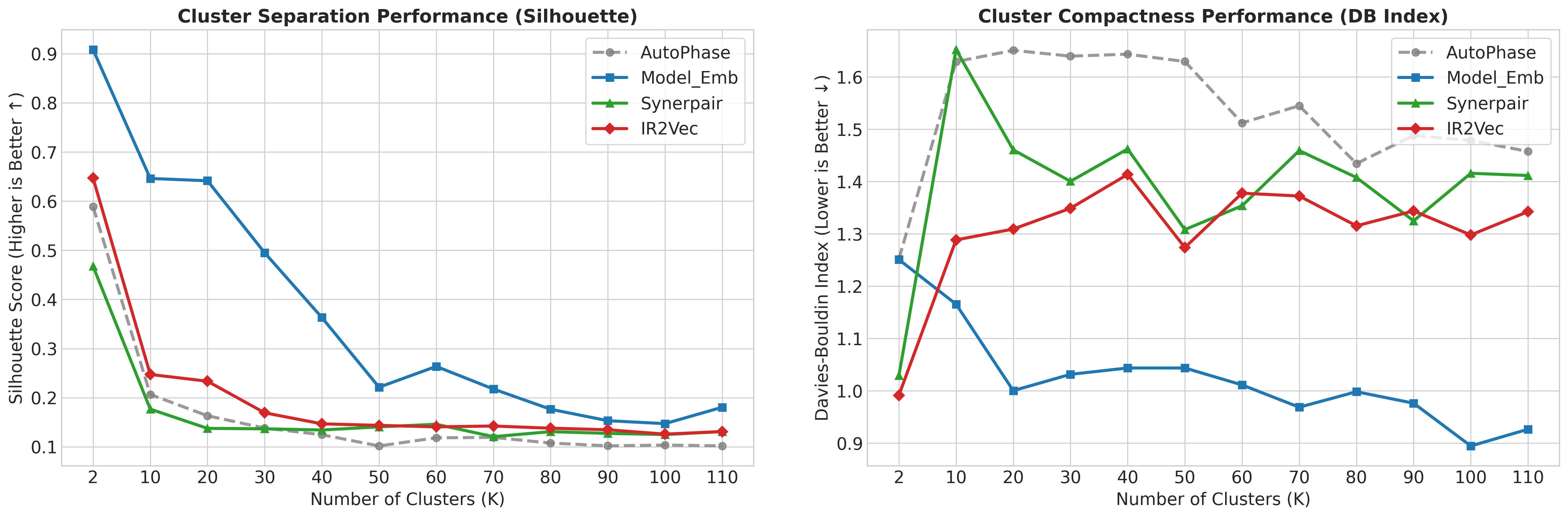} 
    \caption{Comparison of offline clustering quality for different program representations across varying numbers of clusters ($K$). Higher Silhouette Score and lower Davies--Bouldin Index indicate better clustering quality. GRACE's learned embeddings (\textbf{Model\_Emb}) generally produce more compact and better-separated clusters than AutoPhase, Synerpair, and IR2Vec under this evaluation protocol.}
    \label{fig:embedding_quality_comparison}
    \vspace{-1.0em}
\end{figure*}

In addition to cycle-based evaluation, we also measure end-to-end wall-clock runtime on cBench to assess whether cycle-oriented sequence selection translates into practical speedups. The original cBench suite contains 33 programs; 7 were non-functional in our environment, so we report results on the remaining 26 runnable programs. For this experiment, we use the pass sequences obtained under the cycle-reduction objective and measure runtime relative to \texttt{opt -O3}. Following prior work such as PDCAT and CompTuner, we also include \texttt{-O3} in the candidate coreset to improve deployment stability.

Table~\ref{tab:cbench_runtime} shows that the selected sequences outperform \texttt{-O3} on 92.3\% of the reported programs on Intel Xeon 8575C and 84.6\% on AMD EPYC 7763, yielding geometric-mean speedups of 2.7$\times$ and 1.1$\times$, respectively. These results indicate that the cycle-oriented coreset can translate into measurable runtime gains on runnable cBench programs. At the same time, the magnitude of these gains varies across workloads and hardware platforms, so the wall-clock results should be interpreted as a practical validation of the cycle-based objective rather than as a hardware-independent performance guarantee. We also evaluate on PolyBench. The original PolyBench suite contains 30 programs, of which 29 are runnable in our environment. On these benchmarks, the selected sequences achieve arithmetic-mean / geometric-mean speedups of 1.36$\times$ / 1.25$\times$ over \texttt{-O3} on AMD EPYC 7763, and 1.14$\times$ / 1.12$\times$ on Intel Xeon 8575C. This provides further practical support that the cycle-oriented objective can transfer to wall-clock improvements beyond cBench.

\subsection{Generalization Across LLVM Versions}
\label{ssec:llvm18_generalizability}

To examine whether GRACE remains effective beyond the LLVM 10.0.0 setting used for the main baseline comparisons, we conduct an additional evaluation on LLVM 18.1.6. Using the same seven benchmark suites, we measure both IR instruction count reduction relative to \texttt{opt -Oz} and predicted cycle reduction relative to \texttt{opt -O3}. This experiment is intended to assess the robustness of the framework under a more recent compiler release with updated optimization passes and heuristics.

Figure~\ref{fig:llvm18_performance} shows that GRACE remains effective on the newer LLVM release under our evaluation protocol. For code-size optimization, GRACE achieves an average \textit{OverOz} of \textbf{10.3\%}. The gains vary across datasets, with substantial reductions on \texttt{npb} (24.2\%) and \texttt{chstone} (15.3\%), while some benchmarks such as \texttt{mibench} show more modest improvements (2.7\%), possibly because LLVM 18 already provides a stronger baseline. Under the execution-oriented objective, GRACE achieves an average predicted cycle reduction of \textbf{27.5\%} relative to \texttt{opt -O3}. On \texttt{chstone}, the reduction reaches \textbf{53.2\%}, suggesting that the discovered sequences can remain highly effective for some embedded-style workloads under this objective.

These results suggest that GRACE is not tied to a single LLVM release and can remain effective on a newer compiler version as well. At the same time, broader cross-version and cross-compiler validation remains future work.

\subsection{Analysis of Program Embedding}
\label{ssec:embedding_quality}

One component of GRACE is its use of learned program embeddings for offline clustering. To study the role of this component, we first compare the clustering quality of our learned embeddings against several alternative program representations, and then examine how this offline organization affects the subsequent auto-tuning process.

\paragraph{Clustering Quality Evaluation}

To evaluate the quality of the learned representation, we compare GRACE's learned embeddings (\textbf{Model\_Emb}) with three alternative representations: raw \textbf{AutoPhase} features \cite{autophase}, \textbf{Synerpair} features derived from pass-synergy information \cite{cfsat}, and \textbf{IR2Vec} \cite{venkatakeerthy2020ir2vec}. For each representation, we apply k-means clustering over a range of cluster numbers ($K \in \{2,10,20,\dots,110\}$) and assess the resulting partitions using two standard internal validation metrics: the \textbf{Silhouette Score} (higher is better) and the \textbf{Davies--Bouldin Index (DBI)} (lower is better).

The results are shown in Figure~\ref{fig:embedding_quality_comparison}. GRACE's learned embeddings consistently achieve higher Silhouette Scores than the alternative representations across the evaluated cluster numbers, and also obtain the lowest or near-lowest DBI values in most cases. For example, at $K=2$, Model\_Emb achieves a Silhouette Score of 0.91, compared with 0.65 for IR2Vec, 0.59 for AutoPhase, and 0.47 for Synerpair. At $K=100$, Model\_Emb still maintains a Silhouette Score of 0.15, whereas the other representations remain around 0.10--0.13. In terms of DBI, Model\_Emb reaches 0.89 at $K=100$, which is substantially lower than AutoPhase (1.48), Synerpair (1.42), and IR2Vec (1.30). These results suggest that the learned embeddings produce more compact and better-separated clusters under this evaluation protocol.

\paragraph{Impact on End-to-End Auto-Tuning Performance}

\begin{table}[htbp]
\small
\centering
\setlength{\tabcolsep}{3pt} 
\caption{Effect of different program representations on downstream cluster-specific sequence discovery. For each representation, training programs are clustered into 100 groups, and one pass sequence of fixed length 5 is derived for each cluster using deterministic greedy search within the same pass pool $P_{pool}$. The table reports \textit{OverOz} (\%) on LLVM 10.0.0. Higher is better.}
\label{tab:embedding_end_to_end}
\begin{tabular}{lccccc}
\toprule
\textbf{Benchmark} & \textbf{AutoPhase} & \textbf{Random} & \textbf{PairVec} & \textbf{IR2Vec} & \textbf{Model\_emb} \\
\midrule
cbench     & 3.05          & 2.66          & 2.51          & \textbf{3.19}  & 2.65 \\
mibench    & 4.44          & 4.34          & \textbf{4.46} & 4.44           & 4.28 \\
blas       & 0.93 & -2.06         & 0.85          & -0.41          & \textbf{1.22} \\
chstone    & 7.13          & 6.67          & \textbf{7.18} & 7.10           & 7.08 \\
opencv     & 2.95 & -0.36         & 2.68          & 1.84           & \textbf{2.96} \\
tensorflow & \textbf{1.32} & -0.28         & 1.22          & 1.20           & \textbf{1.32} \\
npb        & 9.27          & 7.79          & 10.06         & 11.40 & \textbf{11.48} \\
Avg.       & 4.16          & 2.68          & 4.14          & 4.11           & \textbf{4.43} \\
\bottomrule
\end{tabular}
\end{table}

Having established differences in offline clustering quality, we next examine whether these differences affect the downstream sequence-discovery process. To make this comparison deterministic, we replace the evolutionary search in Stage~3 with a greedy search procedure under a fixed setting. Specifically, for each representation method, we first cluster the training programs into 100 groups, and then, without any initial sequence, we construct a fixed-length pass sequence of length 5 for each cluster from scratch via greedy search within the same constrained pass pool $P_{pool}$. The resulting 100 cluster-specific sequences are then evaluated on the test set. This setup isolates the effect of program organization while avoiding additional variance from stochastic search.

Table~\ref{tab:embedding_end_to_end} reports the resulting average \textit{OverOz} on LLVM 10.0.0. Among the compared methods, our learned embeddings achieve the best overall average (\textbf{4.43\%}), outperforming AutoPhase (4.16\%), random clustering (2.68\%), PairVec (4.14\%), and IR2Vec (4.11\%). The gains are not uniform across all datasets, but the learned embeddings remain competitive on most benchmarks and provide the strongest average performance overall.

These results suggest that the quality of offline program organization affects the quality of the cluster-specific reusable sequences discovered afterward. In other words, grouping programs according to optimization-response-guided embeddings leads to a better basis for subsequent sequence discovery than using raw features, random grouping, or alternative program representations.

\subsection{Ablation Study}
\label{ssec:ablation_candidates_passpool}


\paragraph{Effect of Cluster-Specific Evolution}

We first examine whether the cluster-specific evolutionary stage is necessary, or whether the globally constructed candidate sequences alone are already sufficient. To this end, we remove the cluster-evolution stage and directly evaluate the top 100 sequences from the global candidate set on the test benchmarks. Under this setting, the average \textit{OverOz} values are 5.16\% on \texttt{cbench}, 6.06\% on \texttt{mibench}, 3.39\% on \texttt{blas}, 8.59\% on \texttt{chstone}, 3.53\% on \texttt{opencv}, 1.52\% on \texttt{tensorflow}, and 16.58\% on \texttt{npb}.
Compared with the full GRACE results, these numbers are consistently lower across all datasets. This suggests that the global candidate construction stage alone is not sufficient to produce the strongest reusable sequence set. Instead, the subsequent cluster-specific evolution stage plays an important role by specializing reusable sequences to different groups of programs, rather than relying only on globally strong candidates.

\paragraph{Effect of Global Priors from Stage~1}

We next study the contribution of the Stage~1 global priors. Specifically, we remove the candidate initialization set $C_{seq}$ and replace it with random initialization, while also replacing the pass pool with All pass, keeping the same search budget and optimization objective (10 generations, 25 individuals). This ablation tests whether Stage~1 improves the subsequent cluster-specific search by providing a better starting point.
The results show that removing $C_{seq}$ generally weakens the later search stage. Averaged across the seven datasets, random initialization achieves an \textit{OverOz} of 1.90\%, compared with 4.78\% when the search is initialized from $C_{seq}$. It also converges more slowly on average (8.43 vs.\ 3.29 generations). The quality gap is especially clear on \texttt{blas} (-3.76\% vs.\ 1.56\%), \texttt{cbench} (-11.58\% vs.\ 3.84\%), \texttt{chstone} (2.48\% vs.\ 7.32\%), and \texttt{mibench} (0.98\% vs.\ 5.01\%). There are a few exceptions, such as \texttt{npb} (20.12\% vs.\ 13.88\%) and \texttt{tensorflow} (5.02\% vs.\ 1.10\%), where random initialization performs better under this limited search budget. Overall, however, these results suggest that the candidate sequences constructed in Stage~1 provide useful global priors for the subsequent cluster-specific search.

\paragraph{Quality--Cost Trade-off of Bounded Evaluation}

\begin{figure}[htbp]
    \centering
    \setlength{\belowcaptionskip}{0pt}
    \includegraphics[width=\linewidth]{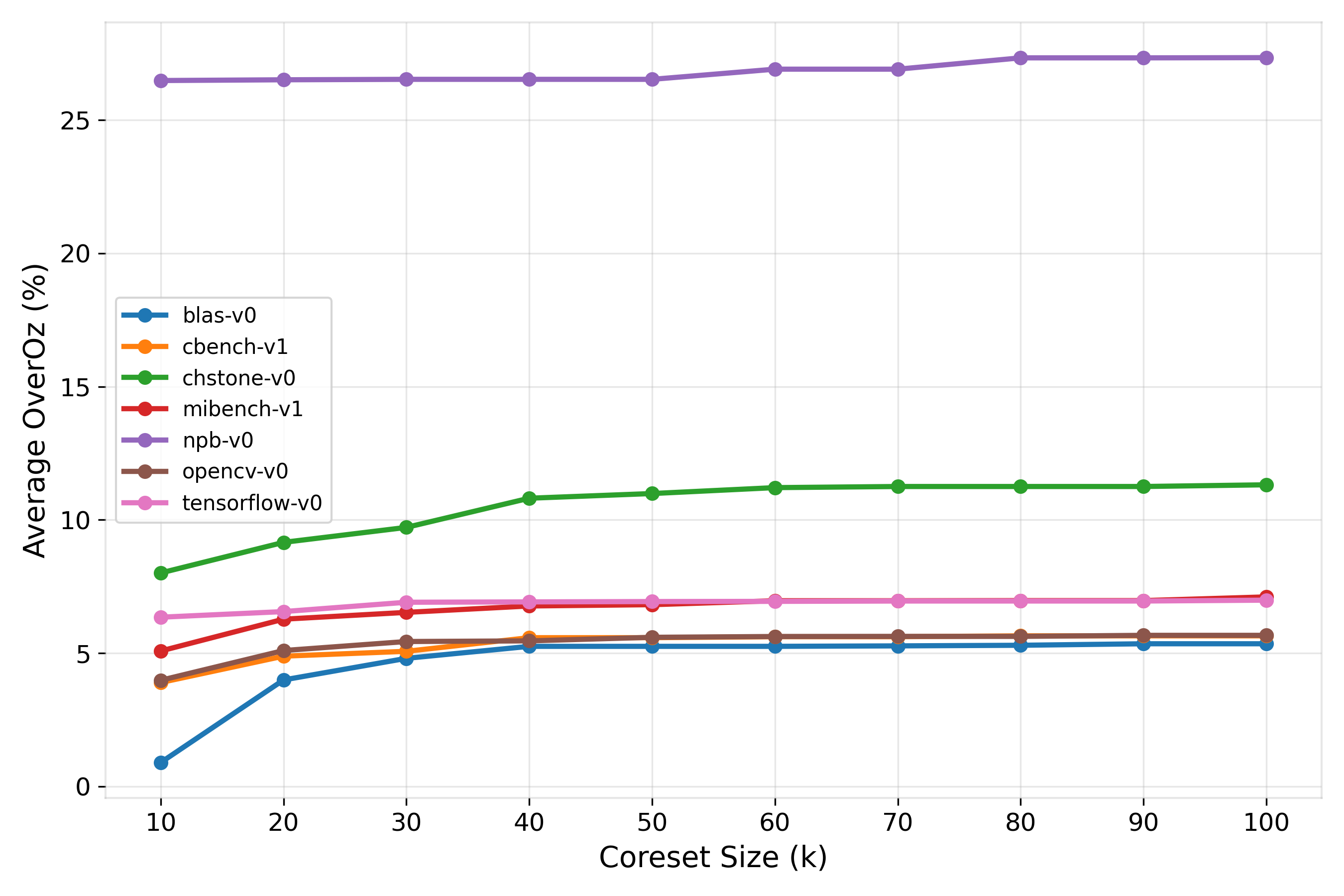}
    \caption{Quality–cost trade-off of bounded coreset evaluation. Larger coresets improve optimization quality but increase test-time cost.}
    \label{fig:k_curve_quality_cost}
\end{figure}

A key design choice in GRACE is to evaluate only a bounded coreset at deployment time. To study the resulting quality--cost trade-off, we vary the coreset size $k$ from 10 to 100 while keeping the rest of the deployment pipeline fixed. For each dataset, we measure the average \textit{OverOz} obtained by evaluating the first $k$ cluster-specific sequences.  As illustrated in Figure~\ref{fig:k_curve_quality_cost}, increasing $k$ improves average \textit{OverOz}, especially in the low-budget regime, while the marginal gains gradually diminish as more sequences are added. In several datasets, a substantial portion of the final improvement is already achieved with moderate coreset sizes, and further enlarging the coreset yields relatively limited benefits. These observations support the bounded-evaluation design of GRACE. While a larger coreset can improve optimization quality, it also incurs higher test-time cost. At the same time, smaller coresets already recover much of the final gain, suggesting that GRACE can be flexibly adapted to different deployment budgets by tuning the coreset size.

\paragraph{Global Sequence Reuse without Clustering}

A more direct comparison to cluster-specific sequence reuse is to remove clustering entirely, run the sequence-evolution stage on the full training corpus, and then deploy the resulting globally evolved sequence set without cluster specialization. Under this setting, the average \textit{OverOz} values are 4.64\% (cbench), 4.77\% (mibench), 1.56\% (blas), 9.59\% (chstone), 2.42\% (opencv), 5.65\% (tensorflow), and 25.06\% (npb). Compared with the full GRACE results, this global sequence-reuse variant performs worse on most datasets. These results suggest that the gains of GRACE are not explained solely by offline sequence evolution itself. Instead, organizing programs into clusters and evolving sequences for each cluster appears to provide a more effective form of reuse than relying on a single globally shared sequence set.

\section{Discussion}
\label{sec:discussion}

The empirical results of GRACE suggest broader observations about compiler auto-tuning beyond its specific implementation.

First, what transfers across programs may not always be a precise \emph{decision function}, but sometimes a \emph{small set of high-quality candidates}. Many learning-based methods aim to predict a single optimization decision or sequence for each unseen program. While appealing in principle, this places strong demands on the learned mapping to generalize accurately across diverse programs. In our experiments, a different strategy appears to work well: distilling a compact set of promising candidate sequences offline and then selecting among them through bounded empirical evaluation. This suggests that, at least in our setting, it may sometimes be more reliable to transfer a reusable candidate set than to rely entirely on predicting a single best sequence.

Second, the results also suggest that reusable pass sequences may not be distributed uniformly across all programs. Some sequences do appear to transfer across programs, but the effectiveness of GRACE indicates that such reuse may be better captured at the level of \emph{program groups} rather than the full training corpus as a whole. One possible interpretation is that different groups of programs share different optimization preferences, so a single globally uniform coreset may be somewhat coarse in this setting. From this perspective, cluster-specific sequence reuse can be viewed as a compromise between purely global reuse and per-program search.

Third, for organizing programs before sequence discovery, optimization response may provide a more useful signal than surface-level similarity alone. In GRACE, positive pairs for representation learning are selected according to similarity in response profiles under several standard optimization levels. This does not imply that the learned embeddings fully characterize optimization behavior, but it does suggest that response-guided program organization can provide a useful basis for subsequent cluster-specific search. More broadly, our results indicate that, for compiler auto-tuning, how programs respond to optimization may sometimes matter more than whether they simply look similar syntactically.

\section{Threats to Validity}
\label{sec:threats}

\paragraph{Internal Validity}
GRACE contains stochastic components, including K-means clustering and genetic search, which may introduce run-to-run variation. To reduce this effect in practice, we constrain the search space using the precomputed candidate set and pass pool, and use fixed experimental settings throughout the evaluation. Still, some variance may remain, and the reported results should be interpreted with this source of randomness in mind. In addition, several design choices and hyperparameters, such as the number of clusters, coreset size, and scoring weights, were selected empirically for the studied setting. Although our empirical results suggest that the framework is reasonably stable within the explored range, different settings may lead to different trade-offs between optimization quality and deployment cost.

\paragraph{Construct Validity}
Our primary code-size metric is LLVM IR instruction count, which is stable and convenient for large-scale evaluation but does not fully capture all aspects of final binary size. For execution-oriented experiments, we use estimated cycle counts from llvm-mca and additionally report wall-clock results on runnable cBench programs. These measurements improve practical relevance, but they are still limited by the chosen platforms, inputs, and measurement conditions. Accordingly, our results should be interpreted as evidence that GRACE is effective under the adopted evaluation protocol, rather than as a complete characterization of downstream hardware behavior in all deployment environments.

\paragraph{External Validity}
As a data-driven framework, GRACE depends on the representativeness of the offline training corpus and may degrade under domain shift or unseen program patterns. We mitigate this by evaluating on seven benchmarks with disjoint training and testing sets and across two LLVM versions. However, our study remains centered on LLVM and does not establish generality across other compiler infrastructures or optimization ecosystems. Moreover, while we study both code-size and execution objectives, broader settings such as energy optimization, multi-objective tuning, or highly latency-sensitive scenarios remain outside this work.

\section{Conclusion}
\label{sec:conclusion}

GRACE is a compiler auto-tuning framework that exploits \emph{cluster-specific sequence reuse} through compact pass-sequence coresets. Instead of predicting a single sequence for each unseen program, it combines global pass-synergy analysis, optimization-response-guided program organization, and cluster-specific evolutionary search to build small sequence coresets for bounded empirical deployment. Experiments on seven benchmark datasets and two LLVM versions show that GRACE consistently improves over standard optimization levels for code-size reduction with low deployment overhead. Under an execution-oriented objective, GRACE also achieves competitive estimated cycle-count reductions and measurable wall-clock speedups on runnable programs. These results suggest that offline-constructed, cluster-specific sequence coresets offer a practical balance between optimization quality and deployment cost.

GRACE also has several limitations. First, its effectiveness depends on the coverage and representativeness of the offline training corpus, since coreset quality is shaped by the programs seen during offline tuning. Second, although deployment cost is bounded and predictable, GRACE still relies on limited test-time empirical evaluation rather than zero-cost sequence selection. Third, our study focuses on LLVM, mainly for code-size optimization, with an additional execution-oriented evaluation under the same framework. Fourth, the optimization-response profiles used for program organization are derived from a small set of standard optimization levels, providing an efficient but coarse signal that richer probing could refine at higher offline cost. Fifth, the number of clusters creates a trade-off between specialization and deployment cost: finer partitioning may improve reuse but enlarge the coreset, while coarser partitioning may weaken specialization. Extending GRACE to broader compiler settings, richer deployment scenarios, and multi-objective optimization remains important future work.


\bibliographystyle{ACM-Reference-Format}
\bibliography{Main}

\end{document}